# Business Intelligence: A Rapidly Growing Option through Web Mining

Priyanka Rahi
(Department of Computer Science
H.P. University, Shimla, H.P., India
Email id: rahi300priyanka@yahoo.com)

**ABSTRACT**
**The World Wide Web is a popular and interactive medium to distribute information in this scenario. The web is huge, diverse, ever changing, widely disseminated global information service center. We are familiar with terms like e-commerce, e-governance, e-market, e-finance, e-learning, e-banking etc. for an organization it is new challenge to maintain direct contact with customers because of the rapid growth in e-commerce, e-publishing and electronic service delivery. To deal with this there is need of intelligent marketing strategies and CRM (customer relationship management) i.e. the effective way of integrating enterprise applications in real time.**
**Web mining is the vast field that helps to understand various concepts of different fields. Web usage mining techniques are attempted to reason about different materialized issues of Business Intelligence which include marketing expertise as domain knowledge and are specifically designed for electronic commerce purposes. To this end, the chapter provides an introduction to the field of Web mining and examines existing as well as potential Web mining applications applicable for different business function, like marketing, human resources, and fiscal administration. Suggestions for improving information technology infrastructure are made, which can help businesses interested in Web mining hit the ground running.**

## 1. INTRODUCTION

The Internet has changed the rules for today's businesses, which now increasingly face the challenge of improving and sustaining performance throughout the enterprise. The growth of the World Wide Web and enabling technologies has made data collection, data exchange and information exchange easier and has resulted in speeding up of most major business functions. Delays in retail, manufacturing, shipping, and customer service processes are no longer accepted as necessary evils, and firms improving upon these (and other) critical functions have an edge in their battle of margins. Technology has been brought to bear on myriad business processes and affected massive change in the form of automation, tracking, and communications, but many of the most profound changes are yet to come.

Leaps in computational power have enabled businesses to collect and process large amounts of data. The availability of data and the necessary computational resources, together with the potential of data mining, has shown great promise in having a transformational effect on the way businesses perform their work. Well-known successes of companies such as Amazon.com have provided evidence to that end. By leveraging large repositories of data collected by corporations, data mining techniques and methods offer unprecedented opportunities in understanding business processes and in predicting future behavior. With the Web serving as the realm of many of today's businesses, firms can improve their ability to know when and what customers want by understanding customer behavior, find bottlenecks in internal processes, and better anticipate industry trends.

This chapter examines past success stories, the current efforts, and future directions of 'Web mining' as an application for business computing. Web usage mining is a part of Business Intelligence rather than the technical aspect. It is used for detecting business strategies through the efficient use of web applications. It is also crucial for customer relationship management (CRM) as it can ensure customer satisfaction as far as the interaction between the customer and organization is concerned. Examples are given in different business aspects, such as product recommendations, fraud detection, process mining, inventory management, and how the use of Web mining will enable growth revenue, minimize costs, and enhance strategic vision.

## 2. WEB MINING

Web mining is a research is converging area from several research communities such as Database, Information Retrieval, Machine Learning and Natural

Language Processing. It is related to the Data Mining but not equivalent to it. Besides a large amount of content information stored on web pages, web pages also contain a rich and dynamic collection of hyperlink information. In addition, web page access and usage information are also recorded in web logs. Web mining [Kosala and Blockeel, 2000] is the use of data mining techniques to automatically discover and extract useful information from web documents and pages. This extracted information enables an individual to promote business understanding marketing dynamics; current trends opted by companies for better growth results etc.

**Web Mining Subtasks**
**Resource Finding:**
The task of retrieving intended web document from the web.
**Information Selection & Preprocessing:**
Automatically selecting and preprocessing specific information from retrieved web resources. This step is transformation process retrieved in IR process from original data. These transformations covers removing stop words, finding phrases in the training corpus, transforming the representation to relational or first order logic form etc.
**Generalization:**
Automatically discovers general patterns at individual website as well as across multiple sites. Data mining techniques and machine learning are often used for generalization.
**Analysis:**
Validation and/or interpretation of mined patterns. In information and knowledge discovery process, people play very important role. This is important for validation and/or interpretation in last step.
Web mining tasks [Kosala and Blocked, 2000] are mainly divided into three classes, namely web content mining, web structure mining, and web usage mining.
Web content mining aims to discover useful information from web content or documents. Basically, web content contains textual data, image, audio, video, metadata and hyperlinks. Most of the web content data are unstructured (free texts) or semi-structured data (HTML documents). The goals of web content mining include assisting or improving information finding (e.g. providing search engines), filtering information based on user profiles, modeling data on the Web, and integrating web data for more sophisticated queries. Text mining and multimedia data mining techniques can be used for mining the content in web pages.
Web structure mining discovers the link structure model based on the topology of hyperlinks on the Web. The link structure model can be used for categorizing web pages and computing the similarity measures or relationships between web pages. It is also useful for discovering authoritative web pages, the structure of web pages itself, and the nature of the hierarchy of hyperlinks in the website of a particular domain.
Web usage mining [Srivastava *et al.*, 2000], also known as web log mining, aims to discover interesting and frequent user access patterns from web browsing data that are stored in web server logs, proxy server logs or browser logs. In this research, we focus on investigating web usage mining techniques to provide enhanced web services.

WEB MINING TAXONOMY
Web Mining can be broadly divided into three distinct categories, according to the kinds of data to be mined. As shown in figure1.
2.1.1 Web Content Mining:
Web Content Mining is the process of extracting useful information from the contents of web documents. Content data corresponds to collection of facts a Web page was designed to convey to user. It may consist of text, audio, video, images, or structured records such as lists and tables.
Text mining and its application to web content has been the most widely researched. Research activities in this field also involve using techniques from AI such as Information Retrieval [IR], Natural Language Processing [NLP], Image Processing and computer vision.

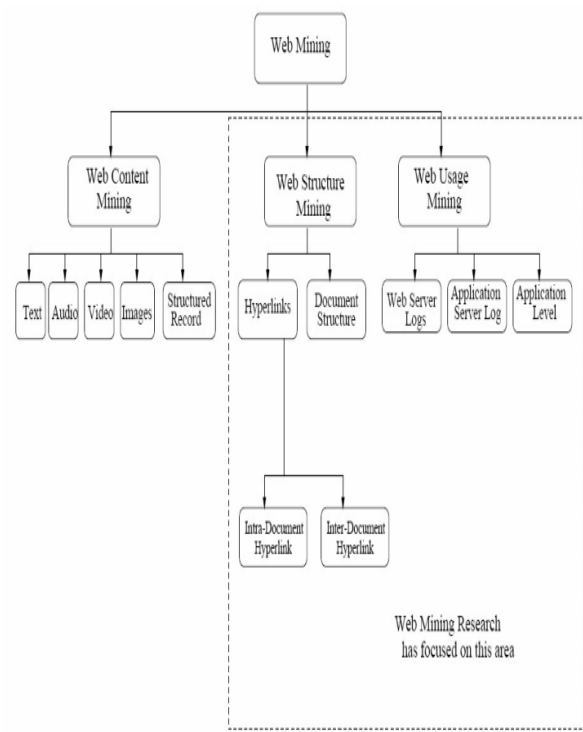

Fig1: *Web Mining Taxonomy [4]*

### 2.1.2 Web Structure Mining :

The structure of typical web graph consists of web pages as nodes and hyperlinks as edges connecting between two related pages. Web Structure Mining can be regarded as the process of discovering structure information from the Web. This type of mining can be further divided into two kinds based on the kind of structural data used.

> Hyperlinks: A Hyperlink is a structural unit that connects a web page to different location either within the same page or to a different web page.
> A hyperlink that connects to a different part of the same page is called <u>Intra Document Hyperlink.</u> And a hyperlink connects two different pages is called an <u>Inter Document Hyperlink</u>. There has been a significant body of work on hyperlink analysis (see survey paper on hyperlink analysis, Desikan et al, 2002).
> Document Structure: The content within a web page can also be organized in a tree structured format, based on the various HTML and XML tags within the page. Mining efforts here have focused on automatically extracting document object model [DOM] structures out of document.

### 2.1.3 Web Usage Mining:

Web usage mining is the application of data mining techniques to identify browsing patterns by analyzing web data i.e. the data residing in web server logs, recording the visits of the users to a website, capturing, modeling and analyzing of behavioral patterns of users in the goal of this web mining category.

## 3. Web Usage Mining

Web usage mining [Srivastava *et al*., 2000] is the application of data mining techniques to discover usage pattern from Web data, in order to understand and better serve the needs of Web-based applications [CMS1997]. Usage data captures the identity or origin of Web users along with their browsing behavior at a web site. Capturing, Modeling and analyzing of behavioral patterns of users is the goal of this web mining category. Web usage mining consists of three phases, namely preprocessing, pattern discovery, and pattern analysis. A high level Web usage mining Process is presented in Figure2 [SCDT2000]. Mobasher et al. [CMS1997] proposes that the web mining process can be divided into two main parts. The first part includes the domain dependent processes of transforming the Web data into suitable transaction form. This includes preprocessing, transaction identification, and data integration components. The second part includes some data mining and pattern matching techniques such as association rule and sequential patterns.

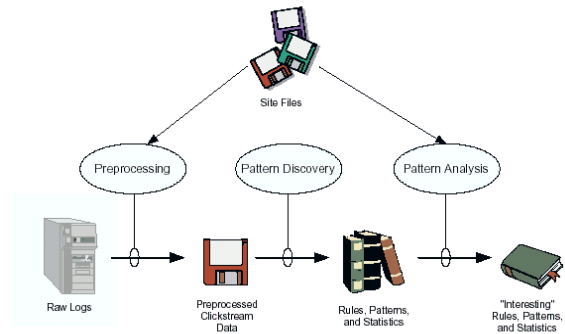

Fig 2: *Web Usage Mining Process*

Web Usage Mining techniques can be used to anticipate the user behavior in real time by comparing the current navigation pattern with typical patterns which were extracted from past Web log. Recommendation systems could be developed to recommend interesting links to products which could be interesting to users. One of the major issues in web log mining is to group all the users" page requests so to clearly identify the paths that users followed during navigation through the web site. The most common approach is to use cookies to track down the sequence of users" page requests or by using some heuristic methods. Session reconstruction is also difficult from proxy server log file data and sometimes not all users" navigation paths can be identified.

### 3.1. Data Sources

The usage data collected at different sources represent the navigation patterns of different segments of the overall web traffic, ranging from single user, single site browsing Behavior to multi-user, multi-site access patterns. Web server log does not accurately contain sufficient information for inferring the behavior at the client side as they relate to the pages served by the web server.

Data may be collected from
- Web servers,
- proxy servers, and
- Web clients.

Web servers collect large amounts of information in their log files Databases are used instead of simple log files to store information so to improve querying of massive log repositories. Internet service providers use proxy server services to improve navigation speed through caching. Collecting navigation data at the proxy level is basically the same as collecting data at the server level but the proxy servers collects data of groups of users accessing groups of web servers. Usage data can be tracked also on the client side by using JavaScript,

Web usage mining itself can be classified further depending on the kind of usage data considered:

Web Server Data: They correspond to the user logs that are collected at Web Server. Some of the typical data collected at a web server include IP addresses, page references, and access time of the users.

Application Server Data: Commercial application servers e.g. Web Logic, Broad Vision, Story Server etc have significant features in the frame to enable E-commerce applications to be built on top of them with little effort. A key feature is the ability to track various kinds of business events and log them in application server logs.

Application Level Data: The new kinds of events can always be defined in an application, and logging can be turned on for them – generating histories of these specially defined events. The usage data can also be split into three different kinds on the basis of the source of its collection:

- On the server side
- On client side
- The proxy side.

The key issue is that on the server side is an aggregated picture of the usage of a service by all users, while on the client side there is complete picture usage of all services by a particular client, with the proxy side being somewhere in the middle.

3.2. Data Pre-Processing

The raw web log data after pre-processing and cleaning could be used for pattern discovery, pattern analysis, web usage statistics, and generating association/ sequential rules. Much work has been performed on extracting various pattern information from web logs and the application of the discovered knowledge range from improving the design and structure of a web site to enabling business organizations to function more efficiently .Data pre-processing involves mundane tasks such as merging multiple server logs into a central location and parsing the log into data fields.

The preprocessing comprises of

- The data cleaning, Consists of removing all the data tracked in Web logs that are useless for mining purposes. Graphic file requests, agent/spider crawling etc. could be easily removed by only looking for HTML files requests. Normalization of URL"s is often required to make the requests consistent.
- The user identification, For analyzing user access behaviors, unique users must be identified. As mentioned earlier, users are treated as anonymous in most web servers. We can simplify user identification to client IP identification. In other words, requests from the same IP address can be regarded as from the same user and put into the same group under that user.
- The Session Identification For logs from a user that spans a long period of time, it is very likely that the user has visited the website more than once. The goal of session identification is to divide weblogs of each user into individual access sessions. The simplest method is to set a timeout threshold. If the difference between the request time of two adjacent records from a user is greater than the timeout threshold, it can be considered that a new access session has started. In this research, we use 30 minutes as the default timeout threshold. and
- The data formatting.

3.3. Pattern Discovery Techniques

Various data mining techniques [Srivastava *et al*., 2000] have been investigated for mining web usage logs. They are statistical analysis, association rule mining, clustering, classification and sequential pattern mining.

3.3.1. Statistical Approach

**Table 3.1** Useful statistical information discovered from web logs.

| Statistics | Detailed Information |
|---|---|
| Web Activity Statistics | Total number of visits<br>Average number of hits<br>Successful/failed/redirected/cached hits<br>Average view time<br>Average length of a path through a site |
| Diagnostic statistics | Server errors<br>Page not found errors |
| Server statistics | Top pages visited<br>Top entry/exit pages<br>Top single access pages |
| Referrers statistics | Top referring sites<br>Top search engines<br>Top search keywords |
| User demographics statistics | Top geographical location<br>Most active countries/cities/organizations |
| Client statistics | Visitor's web browser, operating system, and cookies |

The types of statistical information shown in Table 3.3.1 are usually generated periodically in reports and used by administrators for improving the system performance, facilitating the site modification task, enhancing the security of the system, and providing support for marketing decisions. Many web traffic analysis tools, such as [WebTrends] and [SurfAid], are available for generating web usage statistics.

3.3.2. Association Rule Mining

Association rule mining finds interesting association or correlation relationships among a large set of data items. A typical example of association rule mining is market basket analysis. This process analyzes customer buying habits by finding associations between the different items that customers place in their "shopping baskets". The discovery of such associations can help retailers develop marketing strategies by gaining insight into which items are frequently purchased together by customers. Apriori [Agrawal and Srikant, 1994] is a classical algorithm for mining association rules. Some variations of the Apriori approach for improving the efficiency of the mining process are referred to as Apriori-based mining algorithms. FP-growth [Han *et al.*, 2000] is an efficient approach for mining frequent patterns without candidate generation.

For web usage mining, association rules can be used to find correlations between web pages (or products in an e-commerce website) accessed together during a server session. Such rules indicate the possible relationship between pages that are often viewed together even if they are not directly connected, and can reveal associations between groups of users with specific interests. Apart from being exploited for business applications, the associations can also be used for web recommendation [Lin *et al.*, 2000], personalization [Mobasher *et al.*, 2001]

3.3.3. Clustering

Clustering is a technique for grouping a set of physical or abstract objects into classes of similar objects. A cluster is a collection of data objects that are similar to one another within the same cluster and are dissimilar to the objects in other clusters. A cluster of data objects can be treated collectively as one group in practical applications. There exist a large number of clustering algorithms [Berkhin, 2002]. The choice of a clustering algorithm depends both on the type of data available, and on its purpose and application.

For web usage mining, clustering techniques are mainly used to discover two kinds of useful clusters, namely user clusters and page clusters. User clustering attempts to find groups of users with similar browsing preference and habit, whereas web page clustering aims to discover groups of pages that seem to be conceptually related according to the users' perception. Such knowledge is useful for performing market segmentation in ecommerce and web personalization applications.

3.3.4. Classification

Classification is the process of building a model to classify a class of objects so as to predict the class label of a future object whose class is not known. Since the class label of each training sample is provided, this process is also known as *supervised learning* (i.e., the learning of the model is "supervised" in that it is told to which class each training sample belongs).

For web usage mining, classification is usually used to construct profiles of users belonging to a particular class or category. There is not much work done using classification methods directly for web usage mining due to the complexity of web usage data. In [Tan and Kumar, 2000], it examined the problem of identifying web robot

3.3.5. Sequential Pattern Mining

As mentioned earlier, web logs can be treated as a collection of sequences of access events from one user or session in timestamp ascending order. A web access pattern [Pei *et al.*, 2000] is a sequential pattern in a large set of pieces of web logs, which is pursued frequently by users. Such knowledge can be used for discovering useful user access trends and predicting future visit patterns, which is helpful for pre-fetching documents, recommending web pages, or placing advertisements aimed at certain user groups.

Much research [Srivastava *et al.*, 2000] has been carried out to mine web logs to discover interesting and frequent user access patterns. Sequential pattern mining techniques [Agrawal and Srikant, 1995] are commonly used for discovering web access patterns from web logs.

3.3.6. Dependency Modeling

It is another pattern discovery task in web mining. The goal here is to develop a model capable of representing significant dependencies among various variables in the web domain. As an example, one may interested to build a model representing the different stages a visitor undergoes while shopping in an online store based on the actions chosen (i.e. from casual visitor to a serious potential buyer). There are several probabilistic learning techniques that can be employed to model the browsing behavior of user such techniques include Hidden Markov Model and Bayesian Belief Network. Modeling of web usage pattern will not only provide a theoretical framework for analyzing the behavior of users but is potentially useful for predicting future web resource consumption. Such information may help develop strategies to increase the sales of product offered by the website or improve the navigational convenience of users.

3.4. Pattern Analysis

After discovering patterns from usage data, a further analysis has to be conducted. The exact methodology that should that should be followed depends on the technique previously used. The most common ways of analyzing such patterns are either by using a query mechanism on a database where the results are stored, or by loading the result into a data cube and then performing OLAP operations, visualization techniques, such as graphing patterns or assigning color to different values in the data. Content and Structure information can be used to filter out patterns containing pages that match a certain hyperlink structure.

## 4. HOW WEB MINING CAN AFFECT MAJOR BUSINESS FUNCTIONAL FEATURES

**4. 1.** Web Mining and E-Business correlation

For a number of years AI in the form of Data Mining has been used:

Cellular phone companies, to stop customer attrition.

Financial services firms, for portfolio and risk management.

Credit card companies, to detect fraud & set pricing

Mail catalogers, to life their response rates.

Retailers, for market basket analysis.

Business Intelligence itself is major application area of the Web Usage Mining. In this information on how customers are using a website is critical information for marketers of E-Tailing the business.

This section discusses existing and potential efforts in the application of Web mining techniques to the major functional areas of businesses. These techniques are attempted to reason about different materialized issues of Business Intelligence.

**Table 4.1**. Web mining techniques applicable to different business functions

| Function | Application | Techniques |
|---|---|---|
| Marketing | Product Recommendation, Product Trends | Association Rules Time series data mining |
| Sales Management | Product sales | Multi-stage supervised learning |
| Fiscal management | Fraud detection | Link mining |
| Information Technology | Developer Duplication Reduction | Clustering, Text mining |
| Customer Service | Expert Driven Recommendations | Association Rules, Text mining, Link Analysis |
| Shipping and Inventory | Inventory Management | Clustering, Association Rules, Forecasting |
| Business Process Management | Process Mining | Clustering, Association Rules |
| Human Resources | HR Call Centers | Sequence similarities, Clustering, Association rules |

Some examples that will prove, that the web mining techniques are effectively participated to improve the business functionality. The examples are

[1] Google for mining the web.
[2] Netflix for mining what people would like to rent on DVDs. Which is a DVD recommendation?
[3] Amazon.com for product placement.
[4] Using the eBay Web Services API, developers can create Web-based applications to conduct business with the eBay Platform [Mueller, 2004]. The API can access the data on eBay.com and Half.com. Developers can perform functions such as sales management, item search, and user account management
[6] Web mining techniques can extract knowledge from the behaviour of past users to help future ones, these techniques have much to offer existing e-learning systems.

[7] CiteSeer is one of the most popular online bibliographic indices related to Computer Science. The key contribution of the CiteSeer repository is the ``Autonomous Citation Indexing'' (ACI). Citation indexing makes it possible to extract information about related articles.

[8] Personalisation helps web visitors and customers to find individual solutions in their quest for the content or services that they seek within a web site. The power of the Internet as a two-way channel can be utilised by both the financial service provider and the end user. In terms of the fast emerging area of Customer Relationship Management (CRM), personalisation enables e-business providers to implement strategies to lock-in existing customers, and to win new customers.

## 5. CONCLUSION

We believe that the future of Web mining is entwined with the emerging needs of businesses, Web mining, can aid businesses in gaining an extra information and intelligence. Web mining for business intelligence will be an important research thrust in Web technology— one that makes it possible to fully use the immense information available on the Web. Web usage mining has been gaining a lot of attention because of its potential commercial benefits. Business Intelligence will continue evolving into a more important part of Business Operation as more data from more sources are becoming available and in lower cost. We provide an introduction to Web mining and the various techniques associated with it.

**ACKNOWLEDGEMENTS**

*I would like to express my gratitude to many people. First and foremost, my deepest thanks to my supervisor, Dr. Jawahar Thakur for his invaluable advice and support.*

*I also owe the deepest gratitude to my parents for their deep love. As a daughter, I am really profoundly indebted to them. I hope I will make them proud of my achievements, as I am proud of them.*

*My appreciation also goes to my fellow students and friends in the department, for sharing their knowledge and for many constructive discussions and for their friendship and continuous moral support.*